\documentstyle[12pt]{article}
\begin{document}

\begin{flushright}
EFI 92-29 \\
June 1992 \\

\end{flushright}

\bigskip
\bigskip
\begin{center}
{\bf A HIERARCHIC ARRAY OF INTEGRABLE MODELS\footnote{Work supported in
part by the NSF:  PHY-91-23780}} \\\

\bigskip
\medskip
Peter G. O. Freund \\

{\it Enrico Fermi Institute and Department of Physics \\
University of Chicago, Chicago, IL 60637 } \\

\bigskip
 and \\

 \bigskip
 Anton V. Zabrodin \\

 {\it Institute of Chemical Physics \\
 Kosygina Str. 4, SU-117334, Moscow, Russia} \\

\end{center}
\bigskip
\centerline{ABSTRACT}
\begin{quote}
Motivated by Harish-Chandra theory, we construct, starting from
 a simple CDD\--pole $S$\--matrix,
a hierarchy of new $S$\--matrices involving
ever ``higher'' (in the sense
of Barnes) gamma functions.
These new $S$\--matrices  correspond to scattering of
excitations in ever more complex integrable models.
{}From each of these models, new ones are obtained either by
``$q$\--deformation'', or by considering the Selberg-type
Euler products of which they
represent the ``infinite place''.
A hierarchic array of integrable models is thus obtained.  A remarkable
diagonal link in this array is established.
Though many entries in this array correspond to familiar integrable
models, the array also leads to new models.
In setting up this array we were led to new results on the $q$\--gamma
function and on the $q$\--deformed Bloch\--Wigner function.

\end{quote}

\bigskip
\bigskip
\leftline{\bf 1. INTRODUCTION}
\bigskip

The factorized $S$-matrices, which describe the scattering of
excitations, are known for many $(1+1)$-dimensional integrable models
(e.g. XXZ, XYZ models and their generalizations [1], continuum
sine-Gordon theory [2], etc...).  The Jost functions for the
corresponding input 4-point amplitudes have been identified [3],[4] with
the explicitly calculable Harish-Chandra $c$-functions of certain, in
general {\em quantum} symmetric spaces.

Here we wish to draw attention to unexpected connections between various
integrable models, which arise when certain coupling parameters assume
special {\em rational} values.  These integrable models then naturally
arrange themselves into an ``{\em array of models}''.  By moving from one
site of this array to one of its nearest neighbors, the just-mentioned
connections between known models come into full play.  Sooner or later
though, such moves to nearest neighbors in this ``array of models'' leads
us outside the realm of the known and suggest the existence of new
integrable models.
The hierarchic array is presented in figures 1 and 2, and it leaves, as
we shall see, a number of open questions.
The very existence of this array is our main result.
It implies hierarchical relations between different integrable models,
which look both promising and instructive.

\bigskip
\noindent
{\bf 2.  SINE-GORDON THEORY AT CERTAIN RATIONAL \\
 VALUES OF THE COUPLING CONSTANT}

\bigskip

To begin with, consider soliton\--soliton scattering in the sine\--Gordon
model.
Up to a {\em finite} number of, for us irrelevant factors, the Jost
function for this process is given by the infinite product [2]
$$
A ( u) = \prod_{k=1}^{\infty}
\frac{\Gamma (2k \frac{8 \pi}{\gamma} + i \frac{8 u}{\gamma} )
\Gamma (1+2k \frac{8 \pi}{\gamma} + i \frac{8 u}{\gamma} )}{\Gamma ((2k+1)
\frac{8 \pi}{\gamma} + i \frac{8 u}{\gamma} ) \Gamma (1+(2k-1)
\frac{8 \pi}{\gamma} + i \frac{8 u}{\gamma} )}
\eqno(1a)
$$
where $u$ is the relative rapidity,
$$
\frac{\gamma}{8 \pi} =
\frac{\frac{\beta^2}{8 \pi}}{1 - \frac{\beta^2}{8 \pi}}
\eqno(1b)
$$
and $\beta$ is the sine-Gordon coupling constant.  Now fix
$$
\frac{\beta^2}{8 \pi} = \frac{2}{3} , ~~~~~~~~~~~~~~~ \frac{\gamma}{8 \pi} = 2
{}.
\eqno(2a)
$$
Then
$$
A (u) = K
\frac{\tilde{c}_2 ( \frac{u}{\pi} )}{\tilde{c}_2 (\frac{u}{\pi}+
\frac{1}{2})} ~~,
\eqno(3a)
$$
with $K$ a constant and
$$
\tilde{c}_2 (x) =
\frac{\Gamma_2 (\frac{ix}{2})}{\Gamma_2 (\frac{1}{2}) \Gamma_2
(\frac{ix}{2} + \frac{1}{2})}
\eqno(3b)
$$
where $\Gamma_2$ is the ``higher'' gamma-function [5] related to Barnes'
$G$-function [6] by
$$
\Gamma_2 (x) = (2 \pi)^{-x/2}e^{-x^2/2} G(x-1) ~~.
\eqno(4)
$$
The equations (3) are remarkable on two counts:

	i) the infinite product of eq.~(1a) has disappeared (it is ``hidden'' in
$\Gamma_2$), and

	ii) the function $\tilde{c}_2$ which has appeared, is strongly
reminiscent (see eq. (3b)) of the Harish-Chandra $c$-function [7] of the
real hyperbolic plane \\
(SL(2, {\bf R})/SO(2))
$$
c(x) =
\frac{\Gamma(\frac{ix}{2})}{\Gamma (\frac{1}{2}) \Gamma (\frac{ix}{2} +
\frac{1}{2})}
\eqno(5)
$$
The difference between $c(x)$ and $\tilde{c}_2(x)$ is visible from eqs.~(5)
and (3b):  the gamma function $\Gamma (x)$ is replaced by the ``higher''
gamma function $\Gamma_ 2 (x)$.
A similar $\Gamma \rightarrow \Gamma_2$ transition occurs not only for
$\beta^2 / 8 \pi = 2/3$, Eq.~(2a), but more generally, also for
$$
\frac{\beta^2}{8 \pi} = 1 - \frac{1}{m} , ~~~~
\frac{\gamma}{8 \pi} = m-1 ~~~~m = 4,5,6 \ldots
\eqno(2b)
$$
These are the well-known {\em rational} values of $\beta^2/8 \pi$ at
which the spectrum truncates, as noted by Leclair [2].

The sine-Gordon soliton-soliton $S$-matrix is essentially [4] the $q
\rightarrow 1$ limit of the $S$\--matrix $S_1^{(2)}$ for the scattering
of two (dressed) excitations in the XYZ model.
In terms of the parameter $l$ introduced in [4], the regime (2)
corresponds to the rational values
$$
l = \frac{1}{m-1} ~~~~~~~
m = 3,4,5, \ldots
\eqno(6)
$$
These values of $l$ are special for the transformation [4]
$$
l \rightarrow l ' = \frac{1}{n} \,\,\,\,\,\,\,\,\,\,\,
n \rightarrow n ' = \frac{1}{l} = m - 1
\eqno(7)
$$
which takes one from the magnetic model ${\cal M}_n$
(corresponding to the $Z_n$\--Baxter model) with parameters $ q , l$ to the
${\cal M}_{n '}$ model with parameters $ q , l '$, while leaving the
$S$\--matrix unchanged. It is for and only for the values (6) of $l$ that,
starting from physical values of $l$ and $n$ (i.e. {\em integer} $ n \geq
2$),the transformation (7) yields a new {\em physical} pair of values $l '$
and $n '$ ($n'$ also integer). As will be argued in Section 5 the
symmetric space underlying this argument is the double coseting of a
Kac\--Moody symmetry and the transformation (7) corresponds to the familiar
rank\--level duality [8].  The XYZ-like ${\cal M}_n$ models yield, for
$l$ given by (6), an $S$\--matrix similar to (3), but involving a
$q$\--deformed $\Gamma_2$ function. The situation is similar to
``$S$\--wave'' scattering on ordinary rank one symmetric spaces [9]. There
the Jost function is given [10] by the ratio of two ordinary gamma
functions (eq. (5) above), which when $q$\--deformed is still meaningful:
it corresponds [4] to the scattering of two first\--level excitations in
the $n \rightarrow \infty$ limit of the ${\cal M}_n$ model (the
$R$\--matrix in this limit has been obtained recently by Shibukawa and Ueno
[11]), or in an alternative interpretation [3],[4] to the scattering of
two dressed excitations in the spin $1/2$ XXZ chain.
There is much more to this analogy. To understand this, we first
present some old, as well as some new, results on higher gamma functions.

\bigskip
\noindent
{\bf 3.  ORDINARY AND HIGHER GAMMA FUNCTIONS AND THEIR ${\bf
q}$\--DEFORMATIONS.  THE ${\bf q}$\--BLOCH\--WIGNER FUNCTION}
\bigskip

The Weierstrass product for the ordinary gamma function is [12]:
$$
\Gamma (x) = \gamma (x) \prod_{a=1}^\infty (x+a)^{-1} w_a (x)
\eqno(8a)
$$
with prefactor
$$
\gamma (x) = \frac{1}{x} e^{-Cx} ,
\eqno(8b)
$$
$C = $ Euler-Mascheroni constant, and Weierstrass factors
$$
w_a (x) = ae^{x/a}
\eqno(8c)
$$
The gamma function has half the poles of the function $\pi /\sin \pi x$,
to which it is related by
$$
\Gamma (x) \Gamma (1-x) =
\frac{\pi }{\sin \pi x} = \frac{1}{x} \prod_{a \in Z^*} (x+a)^{-1}a
\eqno(9)
$$
($Z^* =$ set of nonvanishing integers).  $\Gamma (x) $ obeys the
functional equation
$$
\Gamma (x+1) = x \Gamma (x)
\eqno(10)
$$
as well as its higher``multiplication'' generalization due to Gauss.  The
relations (8)-(11) allow for two types of generalization relevant for us.

\bigskip
\centerline{\bf A)  The higher gamma functions}
\bigskip

With the notations
$$
\Gamma (x) = G_1 (x) , \,\,\,\, x = G_0 (x)
\eqno(11)
$$
equation (10) can be written as
$$
G_1 (x+1) = G_0 (x) G_1 (x)
\eqno(12)
$$
which invites the recursive generalization
$$
G_{n+1} (x+1) = G_n (x) G_1 (x)
\eqno(13)
$$
These equations, along with some other requirements (viz. $G_0 (x) = x$ ,
$G_n (1) = 1$, existence of all derivatives of $G_n$ and the condition
that the $(n+1)^{\rm st}$ derivative of $\log G_n (x)$ should be
non\--negative for real $ x \geq 1$ [13]) defines a unique sequence of
meromorphic functions $G_n(x) ~~n = 0, 1, 2, \ldots$ $G_2(x)$ is Barnes'
$G$\--function [6].
Let us call $n$ the {\em height} of the higher $G$\--function $G_n$.
Nowadays it is customary to work not directly with the $G_n$ functions
themselves, but with the {\em higher gamma} functions $\Gamma_n$ which
differ from the $G_n$ by some prefactors [5].  For instance $\Gamma_2
(x)$ is related to $G_2 \equiv G$ via eq.~(4).  We shall also say that
the higher gamma function $\Gamma_n$ has {\em height} $n$.

Like the ordinary gamma functions, the functions $\Gamma_n$ also admit
Weierstrass product representations of the form
$$
\Gamma_n(x) = \gamma_n (x) \prod_{a=1}^\infty (x+a)^{a^{n-1}} w_a^{(n)} (x)
\eqno(14)
$$
with $\gamma_n , w_a^{(n)}$ given in ref.~[5].  Notice that $(\Gamma_n
(x))^{-1}$ has poles at the negative integers.  The pole at $x=-a$ being
of order $a^{n-1}$.  We can rewrite eq.~(15) as an $n$\--fold infinite
product
$$
\Gamma_n(x) = \hat{\gamma}_n (x)
\prod_{a_1=0}^\infty ~
\prod_{a_2=0}^\infty ~
\prod_{a_{n-1}=0}^\infty ~
\prod_{a_n =1}^\infty ~ (x+a_1 + a_2 + \ldots + a_n) \hat{w}_{{a_1}
\ldots a_n}^{(n)}(x)
\eqno(15)
$$
The factors $\hat{\gamma}_n , ~ \hat{w}_{a_1 \ldots a_n}^{(n)}$ can be
found from the references [5], [6].  Yet another form of the infinite
products involves intermediate lower gamma functions instead of $x+a_1 +
\ldots + a_n = \Gamma_0 (x + a_1 + \ldots a_n )$.
For instance, we can write
$$
\Gamma_2(x) = \grave{\gamma}_2 (x) \prod_{a=1}^\infty \left [ \Gamma_1
(x+a) \right ]^{-1} \grave{w}_a^{(2)} (x)
\eqno(16)
$$
More generally $\Gamma_n$ is an $(n-k)$\--fold infinite product of
$\Gamma_k$ functions or their inverses, of course with appropriate
prefactors and Weierstrass factors.

Just as the ordinary gamma function is ``half'' of a trigonometric
function, eq.~(9), so the higher gamma function $\Gamma_2 (x) ,
\Gamma_3 (x) \ldots$ are also related to ``higher'' trigonometric
functions $\Lambda_2(x) , \Lambda_3 (x) , \ldots$  Eq.~(9) generalizes to
$$
\Lambda_n(x) = \Gamma_n(x) \left [ \Gamma_n (-x) \right ]^{(-1)^{n+1}} =
\exp \left
[ - \pi \int_0^x t^{n-1}cot~ \pi t ~ dt \right ]
$$
$$ n = 2, 3, \ldots
\eqno(17)
$$
Notice the alternation of product and ratio of $\Gamma_n$'s in the
definition of $\Lambda_n$.
For $n=1, \Lambda_1 (x)$ is defined [5] as $\pi \exp [- \pi \int_0^x cot
\pi t ~dt ]$ $= \pi (\sin \pi x)^{-1}$ which according to Eq.~(9) is equal
to $\Gamma (x) \Gamma (1-x)$.
The {\em higher trigonometric} function
$\Lambda_n(x)$ is meromorphic of order $n$ and obeys [5]
$$
\log \Lambda_n(x) =
\frac{\Gamma (n)}{(2 \pi i)^{n-1}}
\left [ - \zeta (n) + {\cal D}_n (x) \right]
\eqno(18)
$$
where $\zeta (n)$ is the Riemann zeta function and
$$
{\cal D}_n (x) = \sum_{k=1}^n
\frac{(2 \pi i x )^{n-k}}{(n-k)!} {\rm Li}_k (e^{-i2 \pi x } ) -
\frac{(2 \pi i x)^n}{2n!}
\eqno(19)
$$
is closely related, as noted by Rovinskii [5], to the
Bloch\--Wigner\--Rama\-krish\-nan function
$$
D_n (y) = Re \left[ i^{n+1} \left ( \sum_{k=1}^n \frac{(- \log |y|
)^{n-k}}{(n-k)!}{\rm Li}_k (y) + \frac{(\log |y|)^n}{2n!} \right ) \right ]
\eqno(20)
$$
The function $D_2 (y)$, the original Bloch\--Wigner function, appears in
the calculation of volumes in 3-d hyperbolic (Bolyai\--Lobachewsky)
geometry [15].  The Euler polylogarithm ${\rm Li}_k (x)$, which appears in
eqs.~(19) and (20) is defined as
$$
{\rm Li}_k(x) = \sum_{n=1}^\infty \frac{x^n}{n^k}
\eqno(21)
$$
It will be important for us that the higher gamma functions play the same
r\^{o}le in the functional equation of Selberg zeta functions [16], that
the ordinary gamma functions plays in the functional equation of the
Riemann zeta function [17].  Selberg zeta functions are associated to
double coset spaces $M = \Gamma \backslash G/K$, rank $G/K = 1$, where
$K$ is the maximal compact subgroup of $G$ and $\Gamma$ is a suitable
{\em discrete} subgroup of $G$.  To these data one associates, in the
simplest case of the trivial representation of $\Gamma$, the Selberg zeta
function $Z_M (s)$, defined as usual, by an Euler product.  For instance,
for $G = SO (2,1)$, $K = SO(2)$, this Euler product is [16], [18]
$$
Z_M (z) = \prod_{p \in P} \prod_{k=0}^\infty \left ( 1 - e^{- \tau (p)
(s+k)} \right ).
\eqno(22)
$$
Here $P$ is the set of all {\em primitive} closed geodesics on $M$ (a
closed geodesic is primitive if it is not the multiple traversal of
another closed geodesic) and $\tau (p)$ is the length of the primitive
closed geodesic $p$.  Unlike the prime ideals over which the (simple)
Euler product is taken in the cases of the Riemann and Dedekind zeta
functions, for the Selberg zeta function the product runs over
``primitive elements'' (geodesics) and is a {\em double} product.  Like
the Riemann zeta function, the Selberg zeta function also obeys a
functional equation.  For instance, for $G = SO(2n, 1)$, $K = SO(2n)$
[19], if we define [13], [18], [20]
$$
\hat{Z}_M (s) = Z_\infty (s) Z_M (s)
$$
$$
Z_\infty (s) = \left [ \Gamma_{2n} (s) \Gamma_{2n} (s+1) \right ]
^{(-1)^{n+1} vol M}
\eqno(23)
$$
($Z_M$ as in eq.~(22)), then the functional equation takes the simple form
$$
\hat{Z}_M (2n - 1 - s) = \hat{Z}_M(s).
\eqno(24)
$$
This parallels the Riemann case, where
$$
\hat{\zeta} (s) = \zeta_\infty (s) \zeta (s) ~~~
\zeta_\infty (s) = \pi^{-s/2} \Gamma ( \frac{s}{2} )
\eqno(25)
$$
and the functional equation reads
$$
\hat{\zeta} (1-s) = \hat{\zeta} (s).
\eqno(26)
$$

The second relevant generalization of this ``gamma hierarchy'' concerns
$q$\--deformations.

\bigskip

\noindent
{\bf B) $\bf q$-deformed gamma, higher gamma and Bloch-Wigner functions}
\bigskip

The $q$-deformations of $G_0 (x) = \Gamma_0 (x) = x$ and of $G_1 (x) =
\Gamma_1 (x) = \Gamma (x)$ are classical [21]
$$
\Gamma_{0,q} (x) = [ x]_q = \frac{1-q^x}{1-q}
\eqno(27a)
$$
and
$$
\Gamma_{1,q} (x) = (1-q)^{1-x} \prod_{a=0}^\infty [1 +a]_q [x+a]_q^{-1} .
\eqno(27b)
$$
Notice the simplicity of (27b), as compared to eq.~(8).  The
$q$\--deformation of eq.~(9) we find to be
$$
\Gamma_{1,q}(x) \Gamma_{1,q} (1-x) = i (1-q) (q; q)_\infty^3 q^{{1
\over 8} - {x \over 2}}
\frac{1}{\vartheta_1 (\pi \tau x | \tau )}
\eqno(28a)
$$
where
$$
q = e^{i2 \pi \tau}
$$
$$
(q; q)_\infty = \prod_{a=1}^\infty (1-q^a )
\eqno(28b)
$$
$$
\vartheta_1 (u | \tau ) = 2 q^{1/8} \sin u \prod_{n=1}^\infty
(1-2q^n \cos 2u + q^{2n})(1-q^n) .
$$
In the Jacobi theta function $\vartheta_1$ we have used $q$ related to
$\tau$ as in (28b) rather than the often encountered $\hat{q} = e^{i \pi
\tau}$.  Askey [22] has already pointed out that $\Gamma_{1,q} (x) $
$\Gamma_{1,q}(1-x)$ should involve Jacobi theta functions.  Our equation
(28) makes this fact explicit.  The Jacobi $\vartheta_1$\--function is thus
the $q$\--deformation of the trigonometric sine function.

Higher gamma and trigonometric functions can be similarly $q$\--deformed.
For instance, we define the $q$\--{\em deformed higher trigonometric} function
as
(see eq.~(17))
$$
\Lambda_{2,q} (x) = \prod_{k=1}^\infty
\frac{\Gamma_{1,q}(-x+k)}{\Gamma_{1,q}(x+k)} (1-q)^{-2x}  .
\eqno(29)
$$
The logarithm of $\Lambda_2$ will then correspond to a $q$\--deformed
${\cal D}_2$ function (Eq.~(19)), itself closely related to a
$q$\--deformed Bloch\--Wigner function $D_{2,q}$.
We shall refer to $\log \Lambda_{2,q}$ as a $q$\--deformed Bloch\--Wigner
function, though, strictly speaking, it is ${\cal D}_2$ and {\em not}
$D_2$ whose $q$\--deformation this provides.
A similar $q$-deformed Bloch\--Wigner function was considered in a
different approach by Bloch and Zagier [23].  For the $q$\--analogue of
the $n=2$ case of eqs.~(18), (19) we then readily find
$$
\log \Lambda_{2,q} (x) = \frac{1}{1-q} \int_0^{q^x}
\frac{d_q z}{z} \left \{ \log
 \left [
\frac{iq^{-1/8} z ^{-1/2}}{(q;q)_\infty} \vartheta_1 \left (
\frac{\log z}{2i} | \tau \right ) \right ]\right \}
\eqno(30)
$$
with $\tau, (q; q)_\infty$ and $\vartheta_1$ as in (28b) and $\int d_q z$
standing for Jackson $q$\-- integration [21].  The $q$\--integral of $z^{-1}
\log \vartheta_1$ is thus the $q$\--deformation of the dilogarithmic
expression of eq.~(19).  Similar considerations apply to the
$q$\--deformation of the higher trigonometric functions and of the higher
polylogarithms.

\bigskip
\noindent
{\bf 4.  THE ARRAY OF GAMMA AND ZETA FUNCTIONS AND THEIR SYMMETRIC
SPACES}
\bigskip

With the principals all in place, we can now arrange them in an array
from which their relations become clear.  In this section we present the
array as a mathematical construct.  In the next section we shall show that
to each ``site'' in this array corresponds  a physical scattering
problem with ``geometric'' Jost function. The array is presented in Fig.~1.

The pivotal column of this array is the third column.  Each of the
entries in this column is a gamma function, the height of which increases
in unit steps as we move down the column.  The entry at height one
(second row) is $\Gamma_1$, the building block \`{a} la
Harish\--Chandra\--Bhamamurthy\--Gindikin\--Karpelevi\v{c} of the
$c$\--function of an ordinary symmetric space (see eq.~(5) above and
ref.~[7]).  The function $\Gamma_1$ has this property because $\pi^{-s/2}
\Gamma_1 (s/2)$ is the gamma factor (i.e. the local factor $\zeta_\infty
(s)$ at the infinite place [10], [24]) in the function $\hat{\zeta} (s)$,
(eq.~(25)), in terms of which the functional equation for the Riemann zeta
function takes the simple form (26).

Descending one step down the third column lands us at the function
$\Gamma_2$, itself, as we have seen, the gamma factor for the Selberg zeta
function $Z_M (s)$ of a 2\--dimensional Riemann surface $M_2$.  Further
down at height $2n$ we encounter the gamma factors of Selberg zeta
functions of the $2n$ dimensional domains $M_{2n} = \Gamma \backslash
SO(2n,1)/SO(2n)$. Each of these zeta functions is an Euler product, and
the corresponding Euler factor at primitive element $p$ (at height 1 this
primitive element is a prime number, at height 2 a primitive geodesic,
...) is displayed in the second column.  In the first column we display
the Euler product, gamma factor included, i.e. what we called
$\hat{\zeta} (s)$, $\hat{Z}_M (s)$ in section 2.  Finally, in the fourth
column we present the $q$\--deformation of the gamma function which
appears in the third column at the same height (i.e. in the same row).

The new feature in all this is, that {\em the Euler factor (in the second
column) at height $h$, coincides, up to a constant factor, with the
inverse of the $q$\--deformed gamma function (in the fourth column) at
height $h-1$, provided the deformation parameter $q$ at height $h-1$ is
set equal to the inverse of the ``norm'' of the primitive element at
height} $h$. (By norm of a primitive element we mean the prime number $p$
itself at height $h=1$, $\exp ( \tau (p))$ for a primitive closed
geodesic $p$ of length $\tau (p)$ at height $h=2$, etc...)
This is shown in fig.~1 by the diagonal arrows.
These diagonal arrows express a generalization of the
$p$\--adics\--quantum\--group connection [25], [26], [3], [4].

In the array we only showed explicitly heights 0, 1, 2.  At larger
heights the same pattern is followed, with the remark that for odd
dimensional spaces, the vanishing Euler characteristic leads to a
simplified functional equation [16], [19], [20].

We can now translate all this into statements about integrable models.

\bigskip
\leftline{\bf 5.  THE ARRAY OF INTEGRABLE MODELS}
\bigskip

Starting again at height 1 in the pivotal third column, we encounter the
building block of the Jost function for ``$S$\--wave'' scattering in the
hyperbolic plane $SL(2, {\bf R})/SO(2)$, or equivalently for the
scattering of dressed excitations in the Heisenberg XXX chain.
Next to it (in the second
column) is its adelic partner $\zeta_p (x)$, the building block of the
Jost function
$$
c_p (x) = \frac{\zeta_p (ix)}{\zeta_p (ix +1)}
\eqno(31)
$$
for ``$S$\--wave'' scattering on the $p$\--adic hyperbolic plane $SL(2,
{\bf Q}_p)$/$SL(2, {\bf Z}_p)$ [10],i.e. the Bethe lattice with incidence
number $p+1$.  The Euler product $\hat{\zeta} (x)$ builds the
``$S$\--wave'' scattering on the adelic hyperbolic plane $SL(2, {\bf
A})/$ $SL(2, {\bf Z}_p)$ [10], or the closely related scattering on the
fundamental domain of $SL(2, {\bf Z})$ on the real hyperbolic plane [27].
We see thus that height 1 (second row) in the array corresponds to
$SL(2)$, at least as far as the first three columns are concerned.  The
$q$\--deformation in the fourth column is the Macdonald\--Harish\--Chandra
[25] $c$\--function for root system $A_1$, again $SL(2)$.  This, as we
saw in our earlier work [4], is modeled in the
scattering of two (first level) excitations in the $n \rightarrow \infty$
limit of the ${\cal M}_n$ model, as well as [3] in the
$S$\--wave scattering on quantum hyperboloids, or equivalently, in the
dressed excitation scattering in the spin $1/2$ XXZ model.
So the full height 1 row in the array can be interpreted in
terms of ordinary, or as in the fourth column, quantum symmetric spaces
and modeled on specific physical systems.

Now let us move down to height 2.  In the third column the entry is now a
$\Gamma_2$ function.  As we saw in eqs.~(3) this $\Gamma_2$ builds --
much in
the  way that $\Gamma_1$ did at height 1 -- the Jost function for
soliton\--soliton scattering in the sine\--Gordon model at the values (2)
of the coupling parameter.  What is the symmetric space here?  The
$\Gamma_2$ function is an infinite product of $\Gamma_1$ factors.  This
suggests a Gindikin\--Karpelevi\v{c} type
product [7] over infinitely many roots, therefore a Kac\--Moody
symmetry $\hat{A}_1^k$.
The level $k$ is related to the parameter $l$.
Specifically $k=l^{-1} $.
When eqs.~(2a) or (6) are imposed this yields
$k=m-1$ (remember, $m-1$ is a positive integer).
The transformation (7) then interchanges $l^{-1} = k  =$ level
with $n = $ rank $+1$.
In other words, it becomes {\em rank\--level duality} [8] as already mentioned
in
section 2.
The level $k$ is also related to an
$SU(2)_{\tilde{q}}$ with deformation parameter $\tilde{q}$, a root of
unity $\tilde{q}^{k+2} = 1$ [28].  We are dealing with the $c$\--function
obtained from the {\em zonal} spherical functions of this symmetric
space.  Though the symmetric space itself is infinite\--dimensional, it
is truncated to a single ``radial'' dimension by the {\em double}
coseting involved in going to {\em zonal} spherical functions.

 From this connection with 1-loop Kac\--Moody algebras it also follows
that the height $h$ (of a gamma function) introduced above can be
interpreted as a ``loop number''.  More specifically, the usual symmetric
space of height 1 corresponds to zero loop number, the height 2
Kac\--Moody symmetric space just mentioned to loop number 1 (1\--loop
algebras).  Height 3 is then to be associated with ``2\--loop\--algebras'',
etc... Height
zero corresponds to loop number -1.

The $\Gamma_2$ function, as we saw, can be $q$\--deformed and the
corresponding $\Gamma_{2,q}$ function (fourth column) ``builds'' the Jost
function for the scattering of dressed excitations in the XYZ model, the
true $n=2$ case, for the special values (6) of the parameter $l$ (which
as in ref.~[4] is given by $l = \frac{i \gamma}{\pi \tau}$ with $\gamma$
the anisotopy parameter, and $\tau$ the modular parameter)l.

Moving now to the second column at height two, we find the Euler partners
of $\Gamma_2$ in the Selberg zeta function $\hat{Z}_M$ for which
$\Gamma_2$ provides the gamma factor $Z_\infty$.  Again, the diagonal
connection in the array shows that this yields the
dressed excitation scattering in the XXZ model
(or more generally in the ${\cal M}_\infty$ model) as in the
fourth column at height 1.
Finally, at the ``adelic\--like'' position (first column) at height 2
(third row) we encounter an $S$\--matrix built out of Selberg
zeta\--functions $\hat{Z}_M$ the same way that the adelic $S$\--matrix at
height 1 was built out of $\hat{\zeta}$ functions.  The $S$\--matrix
build out of $\hat{Z}_M$ factors is new and its physical realization is
not known to us.

We can obviously continue this way, down the array and encounter higher
$(n > 2)$ gamma functions and their $q$\--deformations, Euler partners
and $\hat{Z}_M$ functions. They correspond to rank one ordinary or quantum
symmetric spaces of higher dimensionality.  There is the possibility that
the corresponding physical models are also higher dimensional.  This would
be very interesting.

At this point, however, we prefer to take two steps back and move up to
height zero (first row).  Here $\Gamma_0 (x) = x$ is the building block
of a Blaschke\--CDD factor
$$
c_{\rm Blaschke-CDD} (k) = \Gamma_0 (ik-a) = ik-a
\eqno(32)
$$
whence the $S$\--matrix
$$
S_{\rm Blaschke-CDD}= \frac{\Gamma_0 (ik-a)}{\Gamma_0 (-ik-a)}  .
\eqno(33)
$$
By $q$-deforming it we get the type of factor we encountered at height 1
in the second column (``$S$\--wave'' scattering on Bethe lattice).  This
is again the diagonal connection in the array at work.  Just as in
Section 3, the Euler partners and adelic counterparts of the Blaschke
factor, if at all meaningful, are not known to us.  We are now in measure
to produce the {\em array of integrable models}.  It is presented in
Fig.~2.

This array is the main result of this section.  As was emphasized, the
array pivots around its third column.  Moving down this column means
enlarging the symmetric spaces underlying these models.  Moving to the
right (to the fourth column) produces spin chains by $q$\--deformation.
Moving left (to the second column) yields ``Euler partners'' of the
pivotal model of the same height, and finally (first column)
``adelic\--like'' models.  Already at height 2 these are {\em new}
models.  Similarly the height 3, 4, ... models are yet to be physically
identified.  But they are there!

\bigskip
\leftline{\bf 6.  CONCLUSIONS}

\bigskip

We have found here a hierarchical array of gamma functions and, in
effect, of symmetric spaces.  A symmetric space can be $q$\--deformed
(move from the third to the fourth column) or extended while maintaining
rank 1 (move down the ordinary third column, or down the quantum fourth
column).  One can consider its Euler partners (move from the third to the
second column) and its ``adelic\--like'' counterpart (move from the
second to the first column).  To each site in the array we found a
corresponding physical model at special {\em rational} values of some
parameter.  Pushing things to the extreme, we were led to $S$\--matrices
for which the underlying physical model is as yet unknown:  those
involving Selberg zeta functions and those at height $h > 2$.

The mathematical novelties we encountered were:  the $q$\--deformed
Bloch\--Wigner function(eqs.~(29), (30)), the new $q$\--deformed
trigonometric function eq.~(28), and most importantly the array of gamma
functions and symmetric spaces of section 2.

\bigskip
\noindent
{\bf Acknowledgements:}  We wish to thank Spencer Bloch for very valuable
conversations.
A. V. Zabrodin wishes to thank the Mathematical Disciplines Center at the
University of Chicago for its hospitality and support.

\bigskip
\leftline{\bf REFERENCES}

\begin{enumerate}

\item Zamolodchikov, A.B.: Comm. Math. Phys. {\bf 69}, 165 (1979);
Kulish, P.P., and Reshetikhin, N. Yu.: Soviet Phys. JETP {\bf 53}, 108
(1981).

\item Zamolodchikov, A. B., and Zamolodchikov, Al. B.:  Ann. Phys. (N.Y.)
{\bf 120}, 253 (1979); Leclair, A., Phys. Lett.{\bf B230}, 103 (1989).

\item Zabrodin, A. V., Mod. Phys. Lett. {\bf A7}, 441 (1992).

\item Freund, P. G. O., and Zabrodin, A. V.,
Phys. Lett. {\bf 284B}, 283 (1992);  Comm. Math. Phys, in press.

\item Kurokawa, N., Proc. Japan Acad. {\bf 67A}, 61 (1991);
Rovinskii, M. Z., Funct. Analysis and Appl. {\bf 25}, 74 (1991).

\item Barnes, E. W., Quart. Journal Pure and Appl. Math. {\bf 31}, 264 (1900).

\item Helgason, S.: {\em Topics in Harmonic Analysis on Homogenous
Spaces}, Birkh\"{a}user, Basel, 1981.

\item Frenkel, I., Lect. Notes Math. no. 933 Springer, Berlin 1982, p. 71;
Jimbo M., and Miwa, T., Adv. Stud. Pure Math. {\bf 4}, 97 (1984).

\item Olshanetsky, M. A., and Perelomov, A. M.: Phys. Reports {\bf 94},
313 (1983); Wehrhahn, R.F.: Phys. Rev. Lett. {\bf 65}, 1294 (1990).

\item Freund, P. G. O., Phys. Lett. {\bf 257B}, 119 (1991).

\item Shibukawa, Y., and Ueno, K., Waseda preprint, 1992.

\item Whittaker, E. T., and Watson, G. N., {\em A Course of Modern
Analysis}, Cambridge University Press, Cambridge , 14th ed. 1927.

\item Vign\'{e}ras, M. F., Ast\'{e}risque {\bf 61}, 235 (1979).

\item Bloch, S. Lecture Notes, Univ. of Cal., Irvine (1977), unpublished;
D. Ramakrishnan, Contemp. Math. {\bf 55}, 371 (1986);
D. Zagier, Math. Ann. {\bf 286}, 613 (1990).

\item Milnor, J., L'Enseignement Math. {\bf 29}, 281 (1983).

\item Selberg, A., J. Indian Math. Soc. {\bf 20} 47 (1956).

\item Ireland, K., and Rosen, M., {\em A Classical Introduction to Modern
Number Theory}, Springer, N.Y., 1980.

\item Cartier, P., and Voros, A.: in {\em The Grothendieck Festschrift},
vol. 2, Birkh\"{a}user, Basel, 1990, p. 1.

\item Gangolli, R., Illinois J. Math. {\bf 21}, 1 (1977).

\item Kurokawa, N., Contemp. Math. {\bf 83}, 133 (1989).

\item Gasper, G., and Rahman, M., {\em Basic Hypergeometric Series},
Cambridge Univ. Press, Cambridge, 1990.

\item Askey, R., Applicable Analysis {\bf 8}, 125 (1978).

\item Bloch, S. private communication;  Zagier, D., Math. Ann. {\bf286}, 613
(1990).

\item Langlands, R.P,  {\em Euler Products}, Yale Univ. Press, New Haven,
1971.

\item Macdonald, I.G.: in {\em Orthogonal Polynomials:  Theory and
Practice}, P. Nevai ed., Kluwer Academic Publ., Dordrecht, 1990;
Queen Mary College preprint 1989.

\item Freund, P.G.O.: in {\em Superstrings and Particle Theory}, L.
Clavelli and B. Harms eds., World Scientific, Singapore, 1990.

\item Faddeev, L.D., and Pavlov, B. S., Sem. Steklov Math. Inst.
Leningrad {\bf 27}, 161 (1972).

\item Moore, G., and Seiberg, N., Comm. Math. Phys. {\bf 123}, 177 (1989);
Alvarez-Gaum\'{e}, L., Gomez, C., and Sierra, G., Phys. Lett. {\bf 220B},
142 (1989).

\end{enumerate}

\vfill
\eject

\scriptsize

\begin{tabular}{c c c c c c c}
? & $\longleftarrow$ & ? & $\longleftarrow$ & $\Gamma_0(x) = x$ &
$\longrightarrow$ &
$\underline{\Gamma_{0,q}(x) = \frac{1-q^x}{1-q}}$ \\

& & & & & & \\

$\downarrow$ & & $\downarrow$ & & $\downarrow$& &
$\downarrow$ \\

$\downarrow$ & & $\downarrow$ & & $\downarrow$& &
$\downarrow$ \\

& & & & & & \\

$\hat{\zeta}(x) =$ & $\longleftarrow$ & $\zeta_p(x) = $ &
$\longleftarrow $   & $\Gamma_1(x) \sim$ &
$\longrightarrow $ & $\Gamma_{1,q}(x) \sim $ \\

$\zeta_\infty (x) \prod_p \zeta_p(x) $ & & $(1-p^{-x})^{-1}$
 & & $\prod_a (x+a)^{-1}$ &&
$ \underline{\underline{\prod_a (\Gamma_{0,q}(x+a))^{-1}}}$ \\

& &$\underline{\sim [\Gamma_{0,1/p}(x)]^{-1}}$
 & & yields $\zeta_\infty (x)$ & & \\

& & & & & & \\

$\downarrow$ & & $\downarrow$ & & $\downarrow$& &
$\downarrow$ \\

$\downarrow$ & & $\downarrow$ & & $\downarrow$& &
$\downarrow$ \\

& & & & & & \\

$\hat{Z}_M(x) =$ & $\longleftarrow$ & $Z_{M_p}(x) =$& $\longleftarrow$&
$\Gamma_2(x) \sim $& $\longrightarrow$ &
$\Gamma_{1,q}(x) \sim$ \\

$Z_{M \infty}(x) \prod_p Z_{M_p}(x) $& & $ \prod_a (1-e^{- \tau (p)(x+a)})$&
& $\prod_{a,b}(x+a+b)$ & &
$\prod_{a,b} \Gamma_{0,q} (x+a+b)$ \\

 && $\underline{\underline{\sim [\Gamma_{1,e^{-\tau (p)}} (x)]^{-1}}}$
&& yields $Z_{M \infty}(x)$
& &  \\

& & & & & & \\

$\downarrow$ & & $\downarrow$ & & $\downarrow$& &
$\downarrow$ \\

$\downarrow$ & & $\downarrow$ & & $\downarrow$& &
$\downarrow$ \\

\end{tabular}

\medskip

\centerline{\bf Figure 1}

\bigskip

\noindent
NOTE:  The {\em diagonal arrows}, referred to in the text, run
from entries in the fourth column to entries in the second column with
the same number of underlines (i.e., one row below).
These arrows will be displayed explicitly in the preprints mailed out by
ordinary, rather than electronic, mail.

\newpage

\begin{tabular}{c c c c c c c}
 ? & $\longleftarrow$& ? & $\longleftarrow$& Blaschke-CCD Factor &
$\longrightarrow$& ``$S$''-wave scattering \\
& & & & & & \underline{\rm on Bethe lattice} \\
& & & & & & \\

$\downarrow$ & & $\downarrow$ & & $\downarrow$& &
$\downarrow$ \\

$\downarrow$ & & $\downarrow$ & & $\downarrow$& &
$\downarrow$ \\

&&&&&& \\

``$S$''-wave scattering &$\longleftarrow$&``$S$''-wave scattering  &
$\longleftarrow$&
``$S$''-wave scattering &$\longrightarrow$&``$S$''-wave scattering \\

on adelic hyper-& & \underline{on Bethe lattice} && on real hyperbolic
&& on quantum hyper- \\

bolic plane; scat-  &  & & & plane; dressed ex-& & boloid; dressed ex-\\

tering on &&&& citation scattering&&citation scattering \\

fundamental domain &&&& in the XXX model &&
in the XXZ  \\

SL(2,{\bf Z})$\backslash$ SL(2, {\bf R})/SO(2,{\bf R}) &&&&&&
\underline{\underline{and ${\cal M}_\infty$ models} }\\

&&&&&& \\

$\downarrow$ & & $\downarrow$ & & $\downarrow$& &
$\downarrow$ \\

$\downarrow$ & & $\downarrow$ & & $\downarrow$& &
$\downarrow$ \\

&&&&&& \\

? & $\longleftarrow$ & ``$S$''-wave scattering & $\longleftarrow$&
Soliton scattering & $\longrightarrow$ & dressed excita- \\

&& on quantum hyper- & & sine-Gordon model & & tion scattering \\

&& boloid; dressed ex- && at special rational & &tion scattering \\

&& citation scattering & &values of the && in the XYZ model \\

&& in the XXZ  && coupling parameter &&
for $l=1/n$ \\

&&\underline{\underline{and ${\cal M}_\infty$ models}} &&&& \\

&&&&&&\\

$\downarrow$ & & $\downarrow$ & & $\downarrow$& &
$\downarrow$ \\

$\downarrow$ & & $\downarrow$ & & $\downarrow$& &
$\downarrow$ \\

\end{tabular}

\medskip

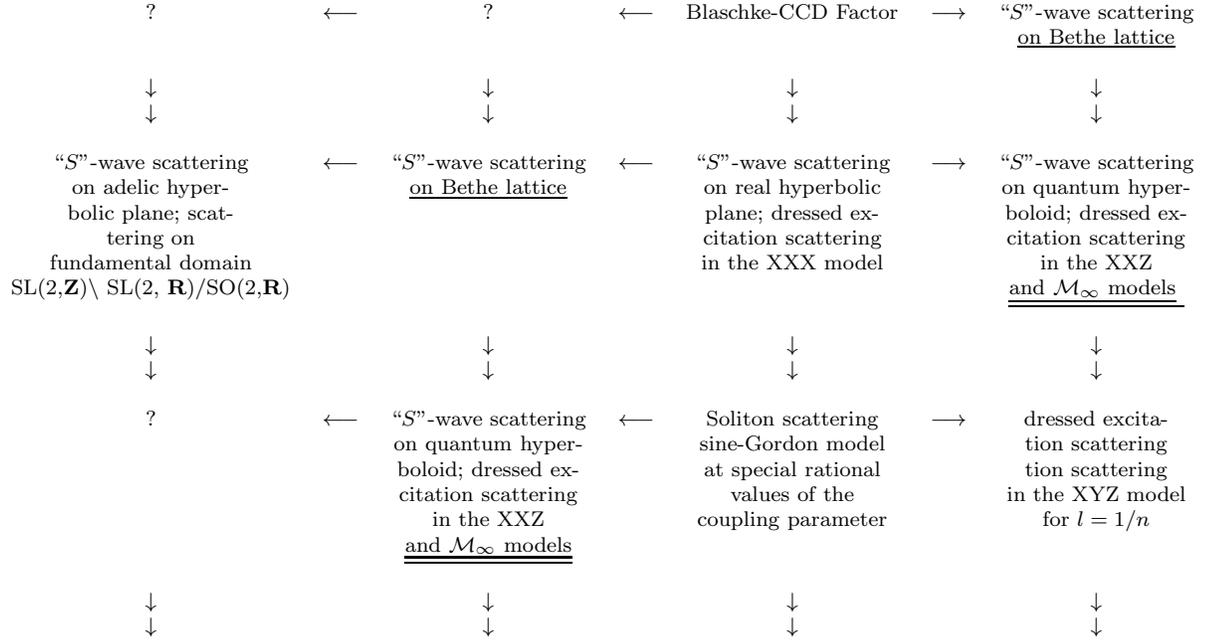
\centerline{\bf Figure 2}

\bigskip
\noindent
NOTE:  The {\em diagonal arrows}, referred to in the text, run
from entries in the fourth column to entries in the second column with
the same number of underlines (i.e., one row below).
These arrows will be displayed explicitly in the preprints mailed out by
ordinary, rather than electronic, mail.

\end{document}